\documentclass[aps,prc,superscriptaddress,showpacs,amssymb,amsmath,amsfonts,twocolumn,floatfix]{revtex4}
\usepackage{graphicx}
\begin{document}

\title{Quantum interference of particles and resonances}

\author{Ya.~Azimov}
\affiliation{Petersburg Nuclear Physics Institute,
             Gatchina, 188300 Russia}

%


\begin{abstract}
Though the phenomenon of quantum-mechanical interference has
been known for many years, it still has many open questions.
The present review discusses specifically how the interference
of resonances may and does work. We collect data on the search
for rare decay modes of well-known resonances that demonstrate
a wide variety of possible different manifestations of
interference. Some special kinds of resonance interference,
not yet sufficiently studied and understood, are also briefly
considered. The interference may give useful experimental
procedures to search for new resonances with arbitrary
quantum numbers, even with exotic ones, and to investigate
their properties.
\end{abstract}

\maketitle

\section{Introduction}

In one of his Physics Lectures, Feynman discussed problems
of scientific imagination. In particular, he asked~\cite{Flec}
`whether it will ever be possible to imagine \textit{beauty}
that we can't \textit{see}. It is an interesting question.
When we look at a rainbow, it looks beautiful to us. Everybody
says, `Ooh, a rainbow.' ... But how would we describe a rainbow
if we were blind? ...  Do we have enough imagination to see in
the spectral curves the same beauty we see when we look directly
at the rainbow?'

Similar problems arise with respect to quantum interference
phenomena. Everybody is sure that the quantum interference does
exist. But one cannot see it directly, only by means of measuring
devices. And it is not always easy to imagine how the interference
will manifest itself in a particular case. As a result, when one
looks at a spectral curve containing interference contributions,
their presence is frequently not recognized (or may be even
refused).

Meanwhile, the interference of resonances has not only academic
interest. Even now it has applications, e.g., to search for and
to study rare decay modes of well-known resonances. Investigation
of $CP$ violation, especially for $B$-mesons, also uses
interference phenomena. The area of their applications may become
wider in future.

In this review, we collect and discuss well-established examples of
interference of resonances. Our aim here is to evolve experience
and intuition of how the interference may work.

\section{General notes: time oscillations of particles}

Everybody knows today that quantum physics, and quantum mechanics
in particular, is probabilistic. However, this appears to be not
its most specific feature. Classical physics may be probabilistic
as well. Even few-body classical systems can reveal a chaotic
time evolution, where the probabilistic description arises quite
naturally. The probabilistic character of statistical physics,
being applied to many-particle systems, is also widely known.

Thus, the basic difference between quantum and classical physics
lies not in probabilities themselves, but in the way the
probabilities should be described for various situations. In
the quantum case, one begins with a wavefunction, its squared
absolute value providing the respective probability. There can
be two (or more) coherent configurations, such that their
wavefunctions may be linearly combined. Then, the resulting
probability contains not only the sum of probabilities for
the two separate configurations, but also an additional term,
describing interference of these configurations. The interference
may be absent by some reason, e.g., if the interfering
configurations are orthogonal. Then the quantum case looks
indistinguishable from the classical one. In the absence of
interference, a physical system could be described in some
classical manner. It is, however, impossible to eliminate
\textit{all} interference contributions from quantum physics,
and just this impossibility enables the Bell inequalities~\cite{BI}
to discriminate between a true quantum case and a hidden-variable
(classical in essence) situation.

One of the most unfamiliar results of quantum physics is the
possibility for some particles to oscillate in time, changing
their characteristics. This may emerge if the corresponding
particles can be transformed to each other, and their
wavefunctions may be coherent.

Of course, the possibility for particles in microworld to
oscillate is a direct manifestation of the particle-wave duality
for quantum objects. The classical notion of a (point) particle
admits the possibility of oscillating motion, but does not admit
oscillation of any internal properties (say, of mass, or some other
characteristic). On the other hand, wave description of a quantum
particle opens different possibilities for classical modeling.
Classical propagation of continuous waves, e.g., sound or
light, is always directly related to some kind of oscillations;
interference of the waves is a well-known and familiar phenomenon.
Note, however, that classical mechanical systems (not a point
particle!), with oscillating motion, may also model quantum
interference effects. A simple example may be given by the
system of two (or more) pendula~\cite{fry}. If they are coupled,
say, by a spring, they move in a correlated manner, which is
reminiscent of the case of interference.

Historically, the first example of particle oscillations was
provided by strangeness mixing and oscillations in the neutral
kaon decays, as suggested by Gell-Mann and Pais~\cite{GMP}. Let
us briefly recall this case (for a more detailed description see
Ref.\cite{fry}; for a brief historical review of the kaon, as well
as neutrino oscillations and quark mixing, see the more recent paper
by Cabibbo~\cite{Cab}; more references, for both experimental data
and their theoretical interpretation, may be found in the Review
of Particle Physics~\cite{PDG08}). Strong and/or electromagnetic
interactions produce neutral kaon states $K^0$ and/or
$\overline{K}^0$, with a definite value of flavor (here it is the
strangeness). However, because of strangeness violation in weak
interactions, definite (and slightly different) eigenvalues of mass
and lifetime belong to other states, $K_S$ and $K_L$, which are
linear combinations of $K^0$ and $\overline{K}^0$. The strangeness
may be tagged by the sign of the electric charge for a lepton,
generated in semi-leptonic decays: $K^0\to l^+$, while
$\overline{K}^0\to l^-$.

If we have initially pure $K^0$ and trace time dependence of the
semi-leptonic decays, then initially we can observe only $l^+$.
However, $K^0$ is a definite coherent combination of $K_S$ and
$K_L$. Since $K_S$ decays more rapidly, the survived combination
of $K_S$ and $K_L$ in the later moments of time will be different
from initial. In terms of flavor, it will inevitably contain
both $K^0$ and $\overline{K}^0$. Therefore, with time, we should
discover generation of both $l^+$ and $l^-$. The ratio of
leptonic yields $l^-/l^+$ changes in time just from the beginning:
it oscillates, tending to a constant ($\approx1$) at large times.
The two lepton yields become here nearly the same, with the
exponential time dependence of pure $K_L$-decays.

Another picture of time evolution is seen if we trace the same
neutral kaon state through pion pairs produced in its decays. The
decay amplitude of $K_L\to 2\pi$ is suppressed by the $CP$-parity
(and is not completely vanishing only due to the $CP$-violation).
Therefore, the $K_L$-contribution is initially very small, and time
dependence of the two-pion yield at small times is almost purely
exponential, corresponding to $K_S$-decays. At later times,
however, the $K_S$-content diminishes, while $K_L$ is dying much
slower. After some time the contributions of $K_S$ and $K_L$ to the
$2\pi$-yield become comparable. Here, they strongly interfere and
provide time oscillations. Even later, the $K_S$-content and its
contribution to the $2\pi$-yield become negligible, so we again
see the pure exponential time dependence, but now characteristic
of $K_L$-decays. Contrary to semi-leptonic decays, the two-pion
channel shows clear non-pure-exponential behavior (and
oscillations) only at some intermediate times, not at early times.

These two examples demonstrate an essential difference between
time dependences of decays for non-interfering and interfering
particles. In a familiar (non-interfering) case, the decay time
dependence is universal, it is the same exponential function for
any possible decay mode. In contrast, oscillations for the
interfering unstable particles are not universal: different
decay channels may have different time dependences.

It is worth also emphasizing that the $K_L$ and $K_S$ have
strongly different mean lifetimes (the ratio
$\tau_L/\tau_S\approx500$~\cite{PDG08}). At first sight, this
could mean that they decay at very different moments of time
and are not able to interfere. Nevertheless, experiments clearly
demonstrate that $K_L$ and $K_S$ can (and do) interfere in
decays! The reason is that the lifetime $\tau$ in quantum
physics is only an average quantity, so the kaon may split
into its decay products at any moment, without waiting its
lifetime. The decay process begins immediately at the moment
of production and continues till the last kaon dies.

Similar flavor oscillations of beauty have been discovered in
decays of neutral $B$-mesons (for both kinds of them, $B_d$
and $B_s$)~\cite{PDG08}. Coherent oscillations of two (or more)
different flavors are also possible~\cite{az}.

Mixing, related to the flavor oscillations, is now known to exist
for neutral $D$-mesons as well~\cite{PDG08}. However, oscillations
themselves cannot be seen in the neutral $D$-meson decays, since
their lifetime is less than 1\% of the oscillation period.

Neutrinos seem to oscillate as well~\cite{PDG08}. More exactly,
only neutrino flavor disappearance has been observed, together
with constancy of the neutrino flux summed over flavors. However,
explicit appearance of a changed flavor, definitely known for
the neutral $K,B$, and $D$ mesons, has not been found yet in
experiments with neutrinos.

Thus, in spite of some uncertainties, existence of the quantum
phenomenon of particle mixing and oscillations is firmly
established. Moreover, such an effect is not unique.

Let us briefly consider the space picture of the particle
oscillations. Again, we begin with neutral kaons. The short-lived
kaons have the mean lifetime $\tau_S=0.9\cdot10^{-10}$~s, which
gives $c\tau_S=2.7$~cm~\cite{PDG08}. Therefore, for realistic
energies, the kaon oscillations take place most probably at
distances of some centimeters or meters from the production
point.  For the $B$-mesons, with $c\tau_B=0.46$~mm~\cite{PDG08},
the oscillations may be seen from some millimeters up to some
centimeters. Contrary to these, oscillation effects for
neutrinos are seen at much larger distances: about $10$~km
for atmospheric neutrinos, some hundreds kilometers for
reactor antineutrinos and accelerator (anti)neutrinos,
and even astronomical distances for solar neutrinos. Thus,
the microscopic phenomenon of quantum interference may
generate quite macroscopic manifestations!

\section{Interference of resonances}

The known hadron resonances have principally the same structure
as the stable hadrons (it is more correct to call them
`stable', since most of them are not really stable, they decay
through weak or electromagnetic interactions). Therefore,
interference of resonances could be considered on the same
ground as, say, interference of neutral kaons. For example, when
studying time evolution of decays into two pions for a coherent
mixture of the meson resonances $\rho^0$ and $\omega$, we should
see a time dependence qualitatively similar to $2\pi$ decays of
a coherent mixture of $K_S$ and $K_L$. Just as $K_L$, $\omega$
is the longer-lived component ($\tau_\omega/\tau_\rho =
\Gamma_\rho/\Gamma_\omega\approx18$~\cite{PDG08}). And, again in
similarity with the $(K_L,K_S)$ case, the amplitude for
$\omega\to2\pi$ is much smaller than for $\rho^0\to2\pi$ (because
of isospin violation, while $K_L\to2\pi$ is suppressed because of
$CP$-violation). Therefore, the time dependence for $2\pi$ decays
of the ($\rho^0, \omega$)-mixture should reveal three regions:
\begin{itemize}
\item Exponential behavior with characteristic time
      $\tau_\rho$, at small times.
\item Manifestations of interference, at some intermediate
      times.
\item Exponential behavior with characteristic time
      $\tau_\omega$, at large times.
\end{itemize}
Regretfully, such a picture is absolutely unobservable. Indeed,
all hadronic resonances decay with very short lifetimes $\tau <
10^{-20}$~s, $~c\tau < 3\cdot10^{-10}$~cm. Therefore, if a
resonance has been produced off a nucleus, the whole space
picture of its decay (including oscillations) sits totally
inside the surrounding atom. Of course, it cannot be seen
(in any sense), and time oscillations for resonance decays,
with such short intervals, cannot be traced.

The situation is, however, not hopeless. While the interference
of resonances cannot be seen as time oscillations, one can use
the complementary variable, the energy (or the mass, in the
rest frame).

It is a frequent opinion that in the energy representation
a resonance reveals itself in the energy distributions only as
a Breit-Wigner (BW) peak of the form
\begin{equation}
    \left|\frac{a}{E-E_0 + i~\Gamma/2}
    \right|^{~2} = \frac{|a|^{~2}}{(E - E_0)^2 + (\Gamma/4)^2}\,.
    \label{q1}
\end{equation}
However, interference, in the energy representation, may violate
such expectations very essentially.

It is interesting to note that the time and energy representations
are not only complementary, but even, with respect to observability
of particle oscillations, appear to be inconsistent. Indeed, as
has just been explained, interference of hadronic resonances is
absolutely invisible in time, but can be studied in the energy
representation (see below). On the contrary, for the neutral
kaons, time oscillations are clearly seen. But if one tried to
study energy distribution for the two pions produced in the kaon
decays, one should see two slightly separated BW peaks,
corresponding to the unstable particles $K_L$ and $K_S$. However,
proper widths of the two peaks and the distance between them are
so tiny ($<10^{-5}$~eV) that could not be measured with any
realistic experimental resolution. For resonances, both peak
widths and distances between peaks are larger, and the peaks may
be separated experimentally. Thus, the mixing and interference of
particles and/or resonances can be studied either in time or in
energy representation, but not in both of them.

Let us return to the interference of resonances. If the energy
dependence of an amplitude contains not only a resonance BW term,
but also some additional contributions, which provide a background
$B$ with respect to the resonance, equation~(\ref{q1}) for the energy
distribution changes and takes the form
    $$\left|B+\frac{a}{E-
    E_0 + i~\Gamma/2}\right|^{~2} =|B|^2 + \frac{|a|^{~2}}{(E - E_0)^2
    + (\Gamma/4)^2} ~~~~~~~~~~~~ $$
\begin{equation}
~~~~~~~~~~~~~~~~~~+ \frac{2|Ba|\cos\varphi\cdot(E - E_0) +
    |Ba|\sin\varphi\cdot\Gamma} {(E - E_0)^2 + (\Gamma/4)^2}\,,
    \label{q2}
\end{equation}
where $\varphi$ is the relative phase between $a$ and $B$. On the
right-hand side of equation~(\ref{q2}), the first two terms provide
the non-coherent sum of the background and BW contributions,
while the third term describes just their interference. Let us
consider properties of the interference in more detail.
\begin{itemize}
\item The interference contribution is linear in both $|a|$
    and $|B|$. Its relative role depends on $|a/B|$. At small
    $|a/B|$ the interference may appear more important than
    the BW contribution itself. This can be considered as
    amplification of a small resonance signal by interference
    with large background.
\item The interference contribution depends on the relative phase
    $\varphi$ between $B$ and $a$.
\item The interference may be either positive (constructive) or
    negative (destructive).
\item In comparison with the BW contribution, the interference
    may have an additional energy dependence and may decrease
    with energy slower than the proper BW contribution.
\item Because of the factor $(E - E_0)$, the interference
    contribution may change its sign. Generally, any BW term
    gives rise to both constructive and destructive interference
    (in different energy regions).
\item The background itself, $|B|$ and $\varphi$, also nay
    depend on energy. As a result, the background may appear
    different in regions of constructive and destructive
    interference, and the relative role of these regions may be
    very different, up to full vanishing of one of them. Thus,
    the presence of interference may provide either bump, or dip,
    or both. Positions of the bump and/or dip are, generally,
    shifted from the true position of the resonance.
\item The same resonance can interfere differently in different
    decay channels, at least due to different properties of the
    corresponding backgrounds.
\end{itemize}

Let us consider now, how the known resonances interfere in
relatively simple cases when two (or more) resonances generate
the same decay products, thus producing the same final states.
Such resonances can still have different quantum numbers, as,
e.g., $\rho^0\to\pi^+\pi^-$ ($J^P = 1^-$) and
$f_0\to\pi^+\pi^-$ ($J^P = 0^+$). At first sight, they cannot
interfere, because of different partial waves of the final
pions ($P$- and $S$-waves, respectively). This is true indeed,
but only after integration over all directions of the relative
momentum of the pions. If one takes a particular direction or
a restricted set of directions, the interference is possible.
It can manifest itself observationally by an asymmetry of the
pions flight direction, in the pair rest-frame, with respect
to a chosen direction (say, to the direction of the laboratory
momentum of the pair). Interference of such a kind seems to be
well understood theoretically. Experimentally, the angular
asymmetry has been recently used to separate the resonances
$\rho^0$ and $f_0(980)$ in photoproduction of the $\pi^+
\pi^-$-pair off the proton~\cite{rho-f}. Note, however,
that the $(\pi^+,\pi^-)$-asymmetry in this reaction may be
influenced not only by the $(S,P)$-wave interference in the
$(\pi^+\pi^-)$-pair, but also by difference in final-state
interactions of $\pi^+p$ and $\pi^-p$.

A clearer case is the interference of direct channel resonances,
e.g., in the annihilation $e^+e^-\to$ hadrons. It may be
called the direct interference of resonances.

\subsection{Resonances and their direct interference in $e^+e^-$
annihilation}

Production of hadrons in $e^+e^-$ annihilation, up to the lowest
order in electroweak interactions, goes through the virtual photon
(or $Z$-boson). Therefore, all direct channel resonances, seen in
the annihilation, have the same quantum numbers $J^{PC}=1^{--}$ as
the photon. Of course, this simplifies theoretical description of
their interference. In addition, experiments on $e^+e^-$
annihilation have reached very high precision of measurements
for some final states.

\begin{figure}[t]
\includegraphics[width=8 cm, keepaspectratio]{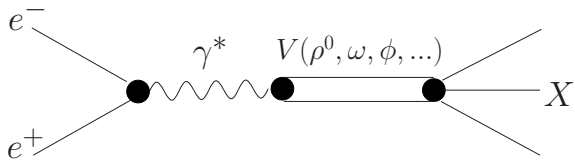}
\caption{The Feynman diagram for $e^+e^-$ annihilation into
         the hadron final state $X$ through a resonance.
         \label{fig:g1}}
\end{figure}

If an isolated resonance decays into the final state $X$ (see
figure~\ref{fig:g1}), its integrated contribution to the process
$e^+e^-\to X$
is proportional to $(\Gamma_{ee}\cdot\Gamma_X)/\Gamma$, where
$\Gamma$ is the total width of the resonance, $\Gamma_{ee}$ and
$\Gamma_X$ are its partial widths for decays into $e^+e^-$ and
$X$, respectively. The energy dependence of the cross section
for an isolated resonance is determined by the BW expression.

If two (or more) resonances in $e^+e^-$-annihilation, with the
same $J^{PC}$, have in addition the same decay products $X$,
they are always coherent and, therefore, can (and, moreover,
should) interfere. Of course, with background determined by
another resonance, equation(\ref{q2}) shows that the interference
becomes weaker (decreases) when the mass difference of the two
resonances increases. Nevertheless, its manifestation may
still be quite essential.

At first sight, two resonances could interfere only if their
BW peaks overlap, i.e., if distance between their masses
is not greater than the sum of their widths. In reality,
however, this is not necessary, since every BW amplitude has
long tails, with not very rapid decrease. Examples below
demonstrate that such a tail may provide sufficient background
to interfere with another resonance.

The situation recalls, to some extent, the time representation
picture of interfering particles with strongly different
lifetimes. For example, in decays of neutral kaons, the $(K_S,
K_L)$-interference becomes to be clearly seen only after several
$\tau_S$, though the $K_S$-component decreases exponentially.
In energy representation, BW tails decrease essentially slower.
As a result (and we will see it below), they provide the
possibility for interference of resonances even if the mass
difference of the resonances is noticeably larger than the sum
of their widths.

Now we are ready to discuss particular cases of direct
interference.

\subsubsection{Final state $\pi^+\pi^-\pi^0$}

\begin{figure}[t]
\includegraphics[width=9.6 cm, keepaspectratio]{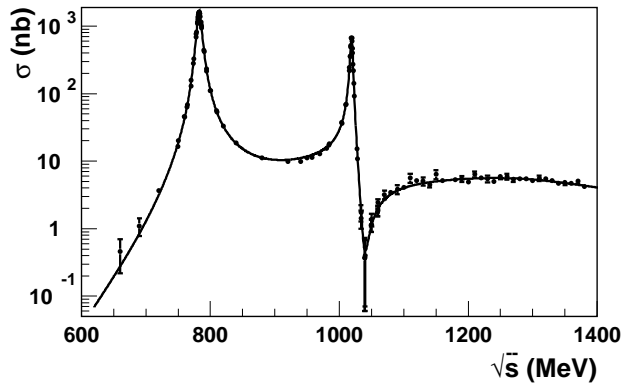}
\caption{The $e^+e^-\to\pi^+\pi^-\pi^0$ cross section measured
     by the SND group and collected in~\protect\cite{SND}.
     The curve is the fit with the $\omega,\phi,\rho^0,\omega'$
     and $\omega''$ resonances. Used with permission of the
     SND group. \label{fig:g2}}
\end{figure}

Experimental data on the reaction $e^+e^-\to\pi^+\pi^-\pi^0$,
as measured by the SND group, are collected in \cite{SND}.
Figure~\ref{fig:g2} shows the cross section as a function of
energy (mass). Its most evident feature is the clear BW peak
of the $\omega$-resonance, which has the mass $m_\omega = 783$~MeV,
the rather narrow total width $\Gamma_\omega = 8.5$~MeV, and the
large branching ratio for the $3\pi$ decay (the partial width
$\Gamma(\omega\to\pi^+\pi^-\pi^0) = 7.6$~MeV). All numerical
values here and below correspond to the latest Review of
Particle Physics~\cite{PDG08}.

The BW tail of $\omega$, together with BW tails of higher
resonances $\omega'$ and $\omega''$, also having essential
decays to the $3\pi$-channel, provides almost constant background
in the $\phi$-meson region, near its tabulated mass $m_\phi =
1019.5$~MeV~\cite{PDG08}. But the $\phi$-meson itself, the
well-known and firmly established resonance, does not evolve
here a standard BW-peak. Instead figure~\ref{fig:g2} demonstrates
nearly ideal interference curve: the cross section shows the
bump, then rapidly drops to dip. After that the cross section
increases again, though slower, up to the value close (but not
equal) to the pre-bump one. The distance between masses of the
maximum and minimum is of order $\Gamma_\phi =4.3$~MeV, and the
$\phi$-meson mass lies just between the values of the maximum and
minimum. Such behavior qualitatively appears to correspond to the
case of equation~(\ref{q2}) with constant background $B$ and small
$|a/B|$ ratio: the interference term is comparable or even larger
than the proper BW term. Indeed, fit to the data leads to the
small partial width $$\Gamma(\phi\to\pi^+\pi^-\pi^0) =
0.65~\textrm{MeV}\,.$$ This is essentially smaller than
$\Gamma(\omega\to\pi^+\pi^-\pi^0)\,,$ in agreement with the
Zweig rule suppression.

The role of the $\rho^0$-meson is worth special consideration.
Masses of $\rho^0$ and $\omega$ are nearly equal (775~MeV and
783~MeV, correspondingly), but the decay $\rho^0\to\pi^+\pi^-
\pi^0$ is expected to be strongly suppressed, due to the isospin
(or, equivalently, $G$-parity) violation. Therefore, at first
sight, the $\rho^0 $ contribution should always be negligible in
comparison with the $\omega$ one. This is true, indeed, near the
vertex of the BW peak, but may be not so for the BW tails,
because of very different total widths. The large $\rho^0
$-width, $\Gamma_{\rho^0}=149.4$~MeV, which is about 18 times
larger than the $\omega$-width, suggests slower decrease of the
$\rho^0$ BW tails as compared to the $\omega$-ones. As a result,
 the $(\rho^0, \omega)$-interference in $e^+e^-\to\pi^+\pi^-
\pi^0$ is quite negligible near the vertex of the BW-peak, but
may be noticeable for BW-tails (below we will see such a situation
explicitly, for the $\pi^0 \gamma$ final state). Fit to
experimental data leads indeed to the very small value~\cite{PDG08}
$$\Gamma(\rho^0\to\pi^+\pi^-\pi^0) = 0.015~\rm{MeV}\,.$$

\subsubsection{Final state $\eta\gamma$}

\begin{figure}[t]
\includegraphics[width=9 cm, keepaspectratio]{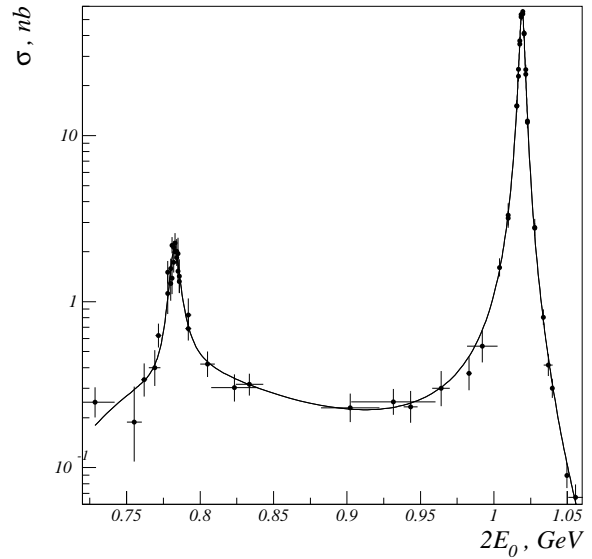}
\caption{The $e^+e^-\to\eta\gamma$ cross section measured by the
    SND group; the curve is the best fit~\protect\cite{SND1}.
    Used with permission of the SND group. \label{fig:g3}}
\end{figure}

Now we consider the reaction $e^+e^-\to\eta\gamma$. Its experimental
cross section, also measured by the SND group, is shown in
figure~\ref{fig:g3}~\cite{SND1}. At first sight, it looks quite natural
and demonstrates two BW peaks, in the $(\rho^0, \omega)$ and $\phi$
regions, without any interference. However, let us consider the
situation in more detail.

We begin with the $(\rho^0, \omega)$ peak. Both $\rho^0$ and
$\omega$ may contribute to the peak, since they both are produced
in $e^+e^-$ annihilation and may decay to $\eta\gamma$. We can
estimate the expected role of the two resonances, since all the
necessary parameters can be determined independently of the
reaction under discussion.

 To the present best knowledge~\cite{PDG08},
\begin{equation}
    \Gamma(\rho^0\to e^+e^-) = 7.04~\textrm{keV}\,,~~~
    \Gamma(\rho^0\to \eta\gamma) = 44.9~\textrm{keV}\,;
    \label{r1}
\end{equation}
\begin{equation}
    \Gamma(\omega\to e^+e^-) = 0.60~\textrm{keV}\,,~~~~
    \Gamma(\omega\to \eta\gamma) = 4.1~\textrm{keV}\,.
    \label{om1}
\end{equation}
Note that the evident smallness of widths (\ref{om1}) as compared
to widths (\ref{r1}) is a result of interference inside the
resonances, at the quark level. In the framework of the
quark-antiquark picture for mesons, both $\rho^0$ and $\omega$
are superpositions of $u\overline{u}$ and $d\overline{d}$ pairs.
For widths (\ref{r1}), the contributions of $u$ and $d$ quarks
interfere constructively and increase the widths, while for widths
(\ref{om1}) their interference is destructive and suppresses
the widths.

The values (\ref{r1}) and (\ref{om1}) suggest that the
$\rho^0$-contribution to the reaction $e^+e^-\to\eta\gamma$ should
dominate over the $\omega$-contribution. Nevertheless, the left peak
in figure~\ref{fig:g3} has its maximum just near the $\omega$-mass,
while near the $\rho^0$-mass one can see only a hint of break. Such
a structure cannot be explained by the sum of two BW peaks, but may
emerge due to constructive interference of the two resonance
contributions. We will discuss this point in more detail below,
in connection with the 2$\pi$ final state.

\begin{figure}[b]
\includegraphics[width=9.8 cm, keepaspectratio]{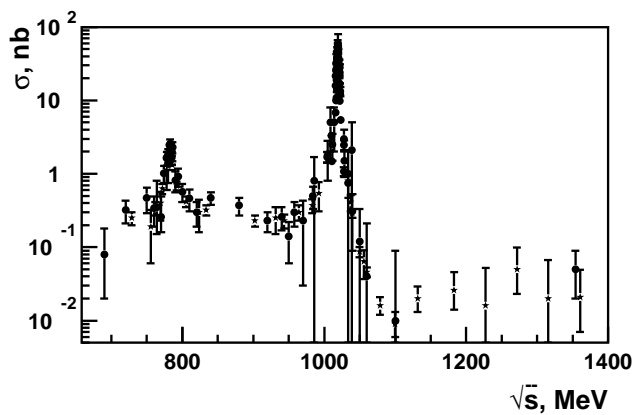}
\caption{The $e^+e^-\to\eta\gamma$ cross section in a wider
    interval of energy~\protect\cite{SND}; the data are
    measurements of the detectors SND (stars) and CMD-2
    (full circles). Used with permission of the SND group.
    \label{fig:g4}}
\end{figure}

The right peak in figure~\ref{fig:g3}, related to the $\phi$-meson,
also gives evidence for the presence of interference: the peak
is not symmetrical, as would be expected for the pure BW term. The right
side of the peak is sharper than the left one. The necessity of the
interference becomes even more evident if one considers the cross
section for $e^+e^-\to\eta\gamma$ in a wider energy interval.
Figure~\ref{fig:g4} shows measurements of detectors
SND and CMD-2, both
below and above the $\phi$-meson~\cite{SND}. The cross sections
just below and just above the narrow $\phi$-peak are
noticeably different. This becomes possible due to the contribution
of the interference term, which decreases, with moving off the
resonance mass, slower than the proper BW contribution, and has
different signs below and above the resonance.

\subsubsection{Final state $\pi^+\pi^-$}

Interference of $\rho^0$ and $\omega$ in decays to $2\pi$ is one
of the earliest and most famous examples of interference of two
resonances.  Moreover, it was the first case where the
interference phenomenon allowed study of a very rare decay,
which could hardly be discovered without interference.

\begin{figure}[t]
\includegraphics[width=8.8 cm, keepaspectratio]{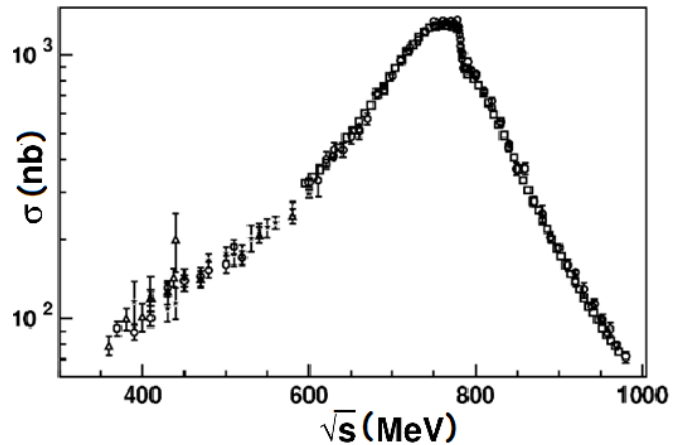}
\caption{The $e^+e^-\to\pi^+\pi^-$ cross section in the region
    of the $\rho^0$ meson peak. The data are measurements of
    the detectors OLYA, CMD, SND, CMD-2 and KLOE (collected
    in~\protect\cite{SND}). Used with permission of the
    SND group. \label{fig:g5}}
\end{figure}

Indeed, the decay $\omega\to\pi^+\pi^-$ is suppressed because of
isospin symmetry violation. Interference contribution of this
decay is also suppressed, but weaker. In addition, contrary to
the proper BW contribution of $\omega$, its interference
with the $\rho^0$-meson is amplified due to the intense
$\rho^0$-production and the nearly 100\% branching ratio for
the decay $\rho^0\to\pi^+\pi^-$. The presence of the $(\rho^0,
\omega)$-interference results in distortion of the canonical
form of the $\rho^0$-meson BW peak. For the reaction $e^+e^-\to
\pi^+\pi^-$, this is well seen in figure~\ref{fig:g5}~\cite{SND},
which shows
\begin{figure}[b]
\includegraphics[width=8.5 cm, keepaspectratio]{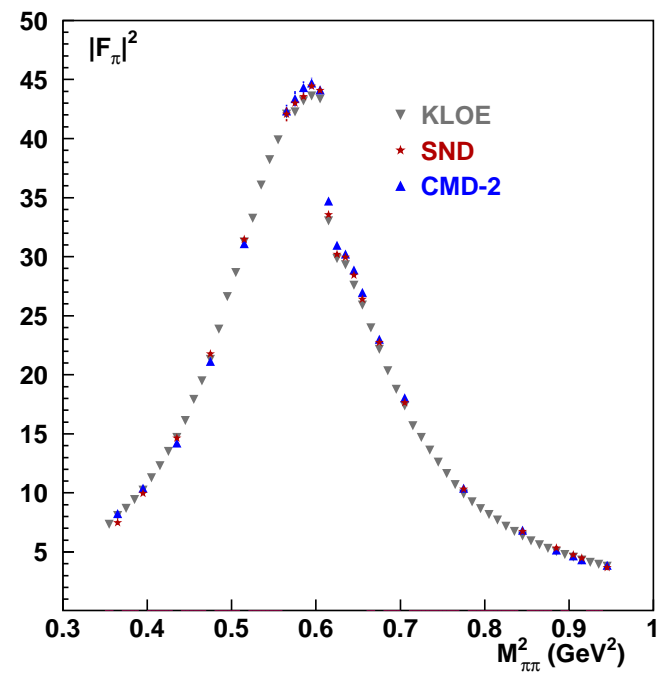}
\caption{The pion form factor squared in the region of
        the $\rho^0$ meson peak. The data are measurements
        of the detectors SND, CMD-2, and KLOE (collected
        in~\protect\cite{KLOE}). Used with permission of
        the KLOE Collaboration. \label{fig:g6}}
\end{figure}
the cross section of this reaction, and/or in
figure~\ref{fig:g6}~\cite{KLOE} for the charged pion
form factor, closely related to the $e^+e^-$ annihilation
into $\pi^+\pi^-$.

The behavior of the form factor may be discussed in more detail
on the basis of measurements by the CMD-2 group~\cite{CMD-2}.
\begin{figure}[t]
\includegraphics[width=8.7 cm, keepaspectratio]{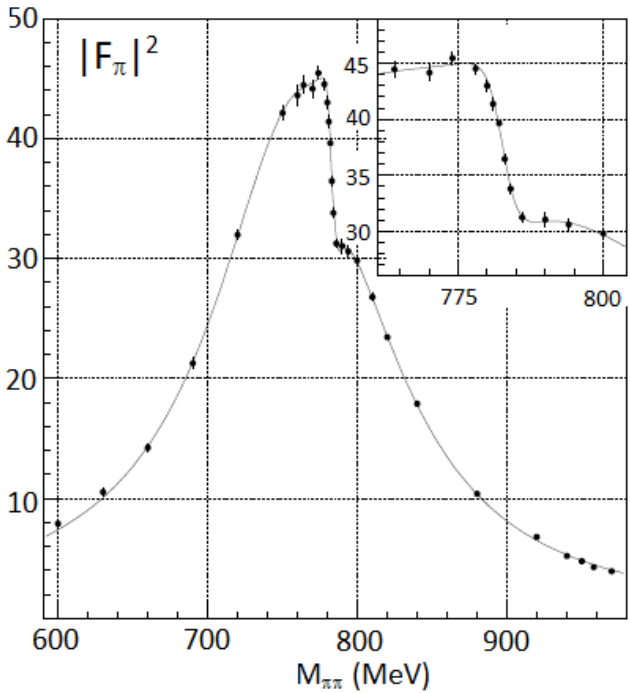}
\caption{The pion form factor squared in the region of the $\rho^0$
    meson peak as measured by the detector CMD-2~\protect\cite{CMD-2};
    the curve is the best fit. The inset shows details of the
    interference region.  Used with permission of the CMD-2 group.
    \label{fig:g7}}
\end{figure}
Their high-precision results (see figure~\ref{fig:g7}) allow us
to consider the inset with details of the interference region.

The left side of the $\rho^0$ peak (below the $m_{\rho^0}$)
looks to be undistorted. But slightly above $m_{\rho^0}$
the form factor rapidly falls due to destructive interference.
If the $\rho^0$- and $\omega$-contributions to the $e^+e^-\to
\pi^+\pi^-$ amplitude were real with respect to each other, then,
according to equation~(\ref{q2}), the interference would change its
sign just at $m_{\omega}$. However, earlier phenomenological
analysis~\cite{az1} presented some evidence for complexity of the
$(\rho^0, \omega)$ mixing. Now we see that the interference changes
its sign somewhat above $m_{\omega}$, thus confirming the presence of
small, but non-vanishing complexity between the $\rho^0$- and
$\omega$-resonance contributions. Parameterization of the fit
shown in figure~\ref{fig:g7}  and the emerging values of the parameters
are given in \cite{CMD-2}. With corrections for the current
$\rho^0$-properties~\cite{PDG08}, they lead to the partial width
\begin{equation}
    \Gamma(\omega\to\pi^+\pi^-) = 0.13~\textrm{MeV}
        \label{ompi}
\end{equation}
(compare with $\Gamma(\rho^0\to\pi^+\pi^-) =
148~\textrm{MeV}\,$~\cite{PDG08}). It is worth emphasizing
that the partial width (\ref{ompi}) is known only due to the
$(\rho^0,\omega)$ interference.

Note that even change of the interference sign, above
$m_{\omega}$, does not lead to the growing behavior of the
form factor anywhere in the interference region. This can be
understood as due to both the mentioned complexity and the
decrease of the interfering background. Indeed, background for
the $\omega$ BW contribution is the $\rho^0$ BW contribution;
in the constructive interference area it is noticeably lower
 than in the destructive one. Therefore, the constructive
interference of $\rho^0$ and $\omega$ is amplified much weaker
than the destructive one, and is hardly seen. Above 800~MeV,
the interference dies out, and the form factor is, again,
determined mainly by the pure $\rho^0$-contribution.

It is interesting to discuss how the picture would look if the
relative sign of the $\rho^0$- and $\omega$-contributions were
just opposite to the existing one. In such a case, we would
need to reverse the above description. Slightly above
$m_{\rho^0}$ the interference would be constructive, and the
form factor would rise, instead of fall. It would look like
some break near $m_{\rho^0}$. Then, about $m_{\omega}$, the
form factor would begin to decrease, partly due to the change of
the interference sign and partly due to the decreasing $\rho^0$
amplitude together with the dying out $\omega$ amplitude. We can
note that, qualitatively, such a hypothetical picture just
corresponds to the structure of the $(\rho^0, \omega)$ peak in
the reaction $e^+e^-\to\eta\gamma$ (see figure~\ref{fig:g3} and
its discussion above).

\subsubsection{Final state $\pi^0\gamma$}

\begin{figure}[b]
\includegraphics[width=8.1 cm, keepaspectratio]{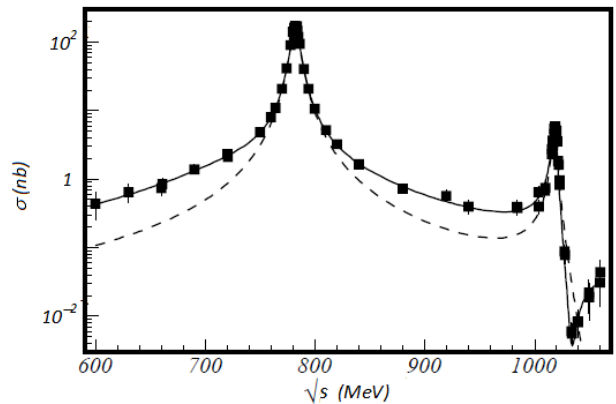}
\caption{The cross section of the process $e^+e^-\to\pi^0\gamma$.
    Data are measurements of the SND group~\protect\cite{SND-03},
    two fitting curves correspond to models with $\rho^0+\omega+\phi$
    (solid) or $\omega+\phi$ (dashed) intermediate states. Used with
    permission of the SND group. \label{fig:g8}}
\end{figure}

The process $e^+e^-\to\pi^0\gamma$ also demonstrates direct
interference of resonances $\rho^0,\,\omega,$ and $\phi$. Its
cross section, measured by the SND group~\cite{SND1,SND-03},
is shown in figure~\ref{fig:g8}. The energy dependence looks here
similar to the case of $3\pi$ annihilation (figure~\ref{fig:g2}).
It has a clear BW-like peak in the $(\rho^0, \omega)$ region
and a bump-dip structure in the $\phi$ region. Again, as in
the $3\pi$ case, for the relative role of the $\rho^0$ and
$\omega$ resonances, we can give qualitative evaluations, based
on other processes and independent of data on the reaction under
discussion.

The decay $\omega\to\pi^0\gamma$ is rather intensive and may be
studied in many reactions.  After all, it has the partial width
(recall that the numerical values correspond to the summary
tables of \cite{PDG08})
\begin{equation}
    \Gamma(\omega\to\pi^0\gamma) = 0.76~\textrm{MeV}\,.
    \label{omg}
\end{equation}
The $\rho^0$ radiative width
\begin{equation}
    \Gamma(\rho^0\to\pi^0\gamma) = 0.09~\textrm{MeV}
    \label{rg}
\end{equation}
is reliably known today just from the $e^+e^-$ annihilation which
we are discussing now. But as its rough estimate one could use the
$\rho^\pm$ radiative width $\Gamma(\rho^\pm\to \pi^\pm\gamma) =
0.067$~MeV, measured in different ways, e.g., by the Coulomb
excitation $\pi^\pm\to\rho^\pm$ (the difference of the two
radiative widths, for $\rho^0$ and $\rho^\pm$, may be explained
by the $(\rho^0,\omega)$ interference, absent for $\rho^\pm$;
see, e.g., Ref.~\cite{az1}).

Note that the essential difference of the radiative widths
(\ref{omg}) and (\ref{rg}) is due to quark-level interference,
just as for the widths (\ref{r1}) and (\ref{om1}). The interference
of $u$ and $d$ quark pairs is constructive (enhancing) for $\omega
\to\pi^0\gamma$ and destructive (suppressing) for $\rho^0\to\pi^0
\gamma$. This difference becomes even stronger in terms of
branching ratios: 9\% and 0.06\% for $\omega$ and $\rho^0$,
correspondingly.

As a result, we can expect that the lower-mass peak in
figure~\ref{fig:g8} is mainly determined by the
$\omega$-contribution. Comparison of the two fits in
figure~\ref{fig:g8}, with and without accounting for the presence
of the $\rho^0$-contribution, confirms such expectation. Near
the peak, the $\rho^0$-contribution is negligible. It becomes
noticeable, however, in the regions aside (below and above) the
$\omega$-peak. The reason was explained above, in connection with
annihilation into the $\pi^+\pi^-\pi^0$ final state. This is mainly
due to BW tails of the $\rho^0$-resonance, which decrease much slower
than BW tails of other resonances, due to the larger total width.

It is interesting to note also that the $\rho^0$-contribution
appears to be practically inessential also near the bump in the
$\phi$-meson region, where the $(\omega, \rho^0)$- and $\phi
$-contributions enhance each other. It becomes essential again
near the dip in the same region, where the $\phi$-contribution
is subtracted from the $(\omega, \rho^0)$-ones.

Concluding the discussion of $e^+e^-$ annihilation into the
final state $\pi^0\gamma$, we see that a good fit without
$\rho^0$-meson is impossible. For $\phi$-meson, the fit
provides a small radiative width $$\Gamma(\phi\to\pi^0\gamma) =
0.005~\textrm{MeV}\,,$$ in accordance with the Zweig rule
suppression.

\subsubsection{Decay $\phi\to\omega\pi^0$}

Of special interest is the decay
\begin{equation}
    \phi\to\omega\pi^0\,.
    \label{phom}
\end{equation}
First, differently to all decays considered above, it is
suppressed twice, since it violates both the Zweig rule and
the isospin symmetry. Therefore, this decay (though possibly
governed by strong interactions, as most resonance decays) is
expected to have a very small partial width (and branching ratio).
Thus, it hardly might be found without interference manifestations.
Second, one of the decay products, $\omega$, is itself a resonance
and, in its turn, can be studied only through its (several) decay
modes. Thus, we need to deal here with a cascade decay, which may
affect the interference picture (in the above examples, only the
$\eta$-meson has similar properties, but it is much narrower than
$\omega$).

The $\omega$-meson has two frequent decay modes, $\,\omega\to
\pi^0\gamma$ (9\%) and $\,\omega\to\pi^+\pi^-\pi^0$ (89\%).
To study decay (\ref{phom}) in $e^+e^-$ annihilation, one may
respectively use processes $$e^+e^-\to\omega\pi^0\to\pi^+\pi^-\pi^0
\pi^0\,,~~~~e^+e^-\to\omega\pi^0\to\pi^0\pi^0\gamma\,.$$ Both
reactions were investigated experimentally in several measurements
of the SND group; their results for the decay (\ref{phom}) have been
 used for the tables of the Review of Particle Physics~\cite{PDG08}
(the corresponding references  are also given there).

\begin{figure}[t]
\includegraphics[width=8.7 cm, keepaspectratio]{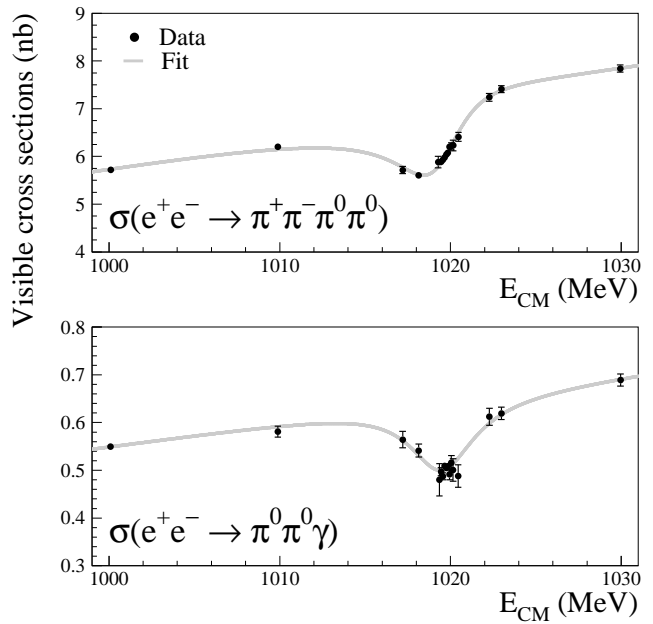}
\caption{The cross sections for the processes $e^+e^-\to\omega
    \pi^0\to\pi^+\pi^-\pi^0\pi^0$ (top) and $e^+e^-\to
    \omega\pi^0\to\pi^0\pi^0\gamma$ (bottom)~\protect\cite{KLOE1}.
    The black dots are the KLOE data and curves are the resulting
     fits. Used with permission of the KLOE Collaboration.
    \label{fig:g9}}
\end{figure}

Recently, the KLOE Collaboration presented new data for the two
reactions~\cite{KLOE1}. They are measured in the same experiment
and with better precision than before. The obtained cross sections
are shown in figure~\ref{fig:g9}, together with the
resulting fits. For each of the cases, the cross section reveals
a dip in the $\phi$-meson region. However, its detailed form is
different for the two cascade branches.

In the bottom panel (decay channel $~\omega\to\pi^0\gamma$), the
lowest point of the dip is at the standard mass value $m_\phi=
1019.5$~MeV, the dip width corresponds to the standard $\phi$-meson
width $\Gamma_\phi=4.3$~MeV. Fits to background before and after the
dip continue each other. The curve in the $\phi$-meson region looks
like the BW peak with the reversed sign. In terms of
equation~(\ref{q2}), such form corresponds to $\cos\varphi=0,~
\sin\varphi=-1$ (i.e., $\varphi=-\pi/2$), and $|Ba|\Gamma > |a|^2$.
Indeed, the accurate fit of the KLOE Collaboration~\cite{KLOE1} gives
the resonance versus background relative phase near $-\pi/2\,$.

The top panel (decay channel $~\omega\to\pi^+\pi^-\pi^0$) shows
different behavior of the cross section. The lowest point of the
dip is reached below $m_\phi$. The background after the dip is
higher than a simple continuation of the background before the
dip. Such properties mean that \mbox{$\cos\varphi>0$} and $\sin
\varphi<0$. These inequalities for the four-pion branch of the
decay cascade are satisfied indeed by the KLOE fit~\cite{KLOE1},
which gives $~\varphi\approx -\pi/4\,.$

Thus, cascade decays of a resonance may provide different
interference pictures in different branches of the cascade, even
with the same first-step decay. It is not amusing, of course,
because the different branches of the cascade generate different
final states, which interfere with different backgrounds.

The KLOE analysis extracts the amplitude for the decay
(\ref{phom}) that corresponds to the branching ratio
$$\textrm{Br}(\phi\to\omega\pi^0)=(4.4\pm0.6)\cdot 10^{-5}\,.$$
It is consistent with the earlier value given in tables~\cite{PDG08},
but is somewhat lower and has twice smaller uncertainty. It leads to
the very small partial width $$\Gamma(\phi\to\omega\pi^0)\approx
0.19~\textrm{keV}\,.$$ Note that this strong-interaction partial
width is smaller indeed than any of the partial widths considered
above, including even radiative widths and widths of decays
into $e^+e^-$. Such is a price of the double suppression
(Zweig rule$\,+\,$isospin symmetry) for the decay (\ref{phom}).

\subsection{Rescattering interference of resonances}

Up to now, we have considered examples of direct interference
of two (or more) resonances. All decay products could come from
decays of any of the two interfering resonances. However, this
is not the only possible form of interference. Resonances
can interfere even if only one of the final particles may be
among decay products of both resonances. Moreover, experimenters
encountered such a kind of interference long ago, in 1960s, when
studying the $\rho$-resonance at the dawn of the era of hadron
resonances.
\begin{figure}[t]
\includegraphics[width=7 cm, keepaspectratio]{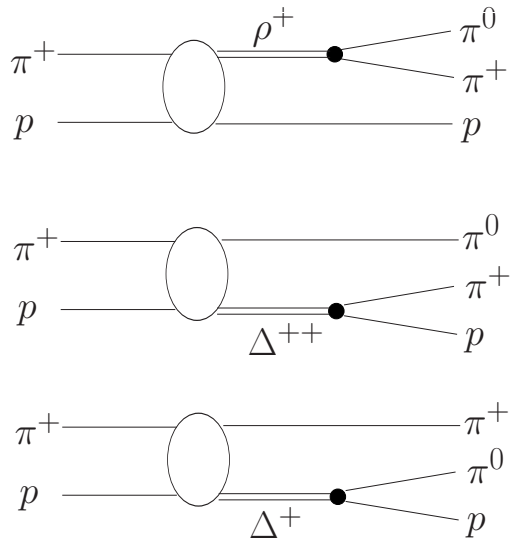}
\caption{Feynman diagrams for possible subprocesses of
        reaction (\protect\ref{pip}). \label{fig:g10}}
\end{figure}

\begin{figure}[b]
\includegraphics[width=7 cm, keepaspectratio]{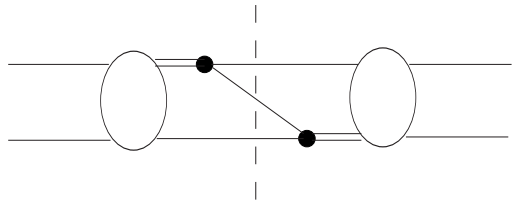}
\caption{A diagram for the cross section contribution of the
    rescattering interference of two resonances.
    \label{fig:g11}}
\end{figure}

\begin{figure}[t]
\includegraphics[width=8 cm, keepaspectratio]{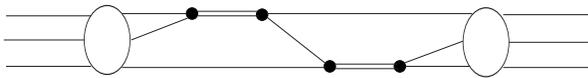}
\caption{Rescattering diagram for three-particle interaction.
        \label{fig:g12}}
\end{figure}

Indeed, the $\rho$-meson may be produced, e.g., in the reaction
\begin{equation}
    \pi^+p\to\pi^+\pi^0p\,,
    \label{pip}
\end{equation}
by the subprocess
\begin{equation}
    \pi^+p\to\rho^+p\to\pi^+\pi^0p\,.
    \label{rop}
\end{equation}
However, reaction (\ref{pip}) may result also from other
subprocesses, e.g.,
\begin{equation}
    \pi^+p\to\pi^0\Delta^{++}\to\pi^+\pi^0p\,,
    \label{dpp}
\end{equation}
\begin{equation}
    \pi^+p\to\pi^+\Delta^{+}\to\pi^+\pi^0p
    \label{dp}
\end{equation}
(Feynman diagrams for the subprocesses (\ref{rop}), (\ref{dpp})
and (\ref{dp}) are shown in figure~\ref{fig:g10}). All the
subprocesses generate the same set of final particles and,
therefore, can interfere. Contribution to the cross section,
provided by such two-resonance interference (figure~\ref{fig:g11}),
has a structure
similar to rescattering diagrams for three-particle
interactions, where a particle interacts first with one
partner, and then proceeds to interact with another one
(figure~\ref{fig:g12}). That is
why interference of such a kind may be called rescattering
interference. Note that the rescattering plays the leading role
in widely used description of three-particle quantum systems by
the Faddeev equations~\cite{feq}.

This kind of interference may also be called the rearrangement
interference, since the observed final particles in this process
rearrange in different ways to reveal different resonances.

The phenomenon of rescattering interference has various analogies
in quantum physics. For instance, in the case of $(\Delta^{++},\,
\rho^+)$ rescattering interference, one cannot discriminate
whether the $\pi^+$ was produced from $\Delta^{++}$ or from
$\rho^+$. This is similar to the case of the two-slit quantum
interference, where one cannot discriminate which of two slits
was traversed by the quantum particle.

However, the cases of direct and rescattering interferences have
essential and interesting differences. Two directly interfering
resonances should possess strictly related quantum numbers: they
both should be either mesons or baryons, they should carry the
same charge and the same flavor - strangeness, beauty, and so on
(we have seen that isospin of two directly interfering resonances
may be different, as for $\rho^0$ and $\omega$; this is possible
since the isospin symmetry is not exact, even in the framework of
strong interactions). In contrast, two resonances providing
rescattering interference may have totally unrelated quantum
numbers. They may be, e.g., a meson and a baryon,
as in reactions (\ref{rop})--(\ref{dp}), they may
even have different charges and/or flavors. All such differences
do not exclude the possibility to interfere.

Instead, the rescattering interference imposes restrictions of
another kind. To be coherent, final states of different processes
(different cascade branches) should, of course, have the same
particle content. But this condition is insufficient. The final
states should also be kinematically consistent. Such consistency
is much more restrictive for rescattering interference than for the
direct one. As a result, the direct interference may be studied as
a function of only one essential parameter, the total energy (total
mass). It is just such consideration that was used above for the
discussion of $e^+e^-$-annihilation. In difference, the rescattering
interference depends on several parameters. For three-particle
production, such parameters are, first of all, the pair masses. The
total energy, at first glance, should not be essential. However, the
kinematical consistency implies that the rescattering interference
of two particular resonances may be noticeable only in a limited
interval of the total energy. Momentum transfers (only two of the
possible six are kinematically independent for transitions of $2\to3$
particles) can also affect the rescattering interference picture.

Existence of the rescattering-type interference, and the
necessity of accounting for it, was clearly demonstrated, e.g.,
by Michael~\cite{mich}. He fitted the $\rho^+$-resonance peak in
reaction (\ref{pip}) at 2.67~GeV/c. The form and parameters of
the peak had been found to vary as a function of position in the
Dalitz plot. The major variations were explained by interference
of the subprocess (\ref{rop}) with other subprocesses. The
dominant effect came from the subprocess (\ref{dpp}). A smaller
contribution was attributed to the production of $\pi N^*$, where
$N^*$ meant nucleon-like $N\pi$ resonances with masses in the
interval $1500-1700$~MeV. It is interesting that the model
description~\cite{mich} has needed also to use interference
with the diffractively produced final states $\pi^+(p\pi^0)$.

Note that the rescattering interference may emerge not only in
particle collisions, but also in particle decays into several
particles. This effect is well known for $B$- and $D$-meson decays.
If the direct interference has become an instrument to study rare
decays of various known meson resonances (see above), the
rescattering-type interference has become an instrument to study
parameters of the $B$- and $D$-meson decays, or relative phases
of final state strong interactions (for a recent example, see
\cite{mes}). In the following subsection, we will discuss
interference effects for decays of heavy hadrons in some more
detail.

Since rescattering interference of two (or more) resonances
(in similarity with direct one) deforms the resonance peaks,
it became a standard approach, when studying resonances, to
eliminate interference as much as possible. For such a purpose,
intensively produced resonances (say, $\Delta^{++}$ in
reaction (\ref{pip}) ) are usually cut out. This may be the
reason why the existing literature does not provide any study
of the general structure and properties for the rescattering
interference, though there are many model-dependent
considerations which fit data on particular reactions through
accounting for rescattering-type interferences of various
resonances.

Elimination of interference is an appropriate method for
extracting a resonance with sufficiently high apparent cross
section of its production, in comparison with non-resonant
contributions. The situation may be different, however, for
rare decay modes, or if there exist resonances with
relatively low production cross section.

Possible existence of unusual hadrons, having suppressed
couplings to the familiar hadrons, as a result of specific
internal structure, was suggested long ago~\cite{az2}. With
such states, interference of some kind could be really helpful
to search for their manifestations.

Indeed, the exotic baryon $\Theta^+$ (initially, $Z^+$) was later
predicted~\cite{dppt} on the basis of the quark-soliton model.
It should have evidently non-canonical quark structure and, by
prediction~\cite{dppt}, is expected to have small width (and
small decay coupling). Its production may represent a new kind
of hard processes~\cite{ags}, thus implying relatively small
production cross section. In any case, $\Theta^+$ has not
been seen in many experiments. If $\Theta^+$ does,
nevertheless, exist, its production is strictly limited.

In particular, rather low upper boundary for the
$\Theta^+$-photoproduction on the proton was given by the
CLAS Collaboration~\cite{clas}. To amplify a possible signal
of the $\Theta^+$, it was suggested~\cite{adp} to look for
the interference of final states for two subprocesses (see
\begin{figure}[t]
\includegraphics[width=7 cm, keepaspectratio]{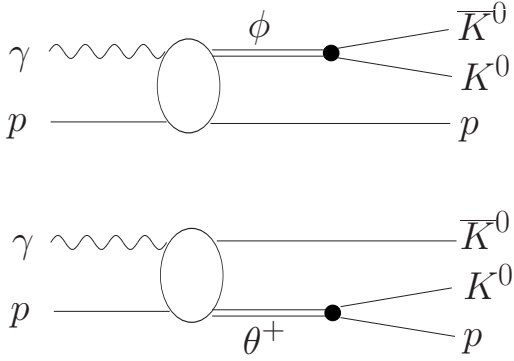}
\caption{Diagrams providing rescattering interference to
    amplify the $\Theta^+$-signal in the $K_S K_L p$
    photoproduction. \label{fig:g13}}
\end{figure}
figure~\ref{fig:g13})
\begin{equation}
    \gamma p\to\phi p\to K_SK_Lp\,,~~~~
    \gamma p\to K_S\Theta^+\to K_SK_Lp\,.
    \label{fitet}
\end{equation}
This is just a rescattering-type interference, similar to the
interference of $\rho$ and $\Delta$ discussed above. But this
time, a possible weak (and unobserved) proper $\Theta^+$-signal
may be enhanced by the strong signal of the $\phi$-resonance.
Note that contribution of the $\phi$-photoproduction is cut
out in the published analysis~\cite{clas}, and any potential
interference with  $\phi$ has been discarded.

\begin{figure}[t]
\includegraphics[width=7 cm, keepaspectratio]{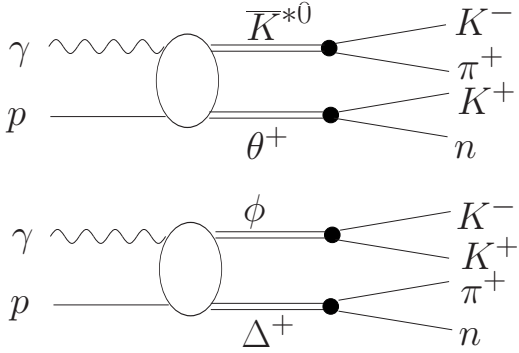}
\caption{Diagrams providing rescattering interference to
    amplify the $\Theta^+$-signal in the $K^-K^+\pi^+ n$
    photoproduction. \label{fig:g14}}
\end{figure}

Subprocesses leading to multiparticle final states with more
than three particles also can (and should) interfere. Such processes
depend on even larger number of physical parameters (pair masses
and momentum transfers), which may affect the possible interference
picture. It is, therefore, essentially more complicated type of
rescattering interference than the three-particle cases discussed up
to now in this subsection. Nevertheless, it can also be helpful to
search for new resonances and investigate them. For instance, for
$\Theta^+$, interference of the subprocesses
$$\gamma p\to\phi\Delta^+\to K^-K^+\pi^+n \,,$$
\begin{equation}
    \gamma p\to\bar K^{*0}\Theta^+\to K^-\pi^+K^+n\,,
    \label{deltet}
\end{equation}
shown in figure~\ref{fig:g14}, is also suggested to be
investigated~\cite{adp}. Here, the two good resonances, $\phi$ and
$\Delta$, may enhance manifestation of $\Theta^+$.

Note an interesting difference between processes (\ref{fitet})
and (\ref{deltet}). The process (\ref{fitet}) may provide a peak
in the system $K_L p$ (or $K_S p$), and one could not discriminate
between $\Sigma$-like state (with $S=-1$) or $\Theta$-like state
(with $S=+1$). In contrast, a peak in the system $K^+ n$, expected
in the process (\ref{deltet}), has the tagged strangeness $S=+1$.

\subsection{Interference phenomena in decays}

As was mentioned in the previous subsection, various kinds of
interference may be seen also in decays of heavy hadrons. They may
be similar to the direct interference, or the rescattering one, or
be of more complicated type (as will be seen below). For example,
an essential part of decays $J/\psi\to\pi^+\pi^-\pi^0$ goes through
the intermediate two-body channels $\rho^0\pi^0, \,\,\rho^+\pi^-,
\,\,\rho^-\pi^+\,$ (see the corresponding branching ratios in the
tables~\cite{PDG08}). Of course, all these channels interfere with
each other (just as in rescattering interference) and, thus, affect
the distribution of events over the Dalitz plot. Therefore, to
accurately extract the coupling constant between $J/\psi$ and
$\rho\pi$, one should account for the interference. Such necessity
becomes even more important for decays of even heavier hadrons.

Collaborations BaBar and Belle, working at $B$-factories, have
collected great sets of data on multiparticle decays of $B$-,
$D$-mesons, and some other heavy hadrons. A large set of data on
$D$-meson decays has been gathered also by the CLEO Collaboration.

Those decays provide many examples of various kinds of interference.
They are worth a special review paper, and we will not consider
here all details of interference in the decays. Instead, we will
mainly be concerned with similarities and/or differences with respect
to the interference manifestations described above. Nevertheless, we
will briefly discuss also some particular examples.

The interference picture in collisions, as in $e^+e^-$ annihilation
or, e.g., in reaction (\ref{pip}), depends on the total energy.
It essentially changes (or even disappears) when the total energy
changes. For decays, in contrast, the total energy is fixed by
the mass of the decaying particle. Moreover, collisions generally
produce states with various values of parity and total angular
momentum. In contrast to this, decays produce only final states
with the $J^P$ value of the initial particle. In this respect, the
situation is similar to the $e^+e^-$ annihilation where hadrons
are produced through the virtual photon with fixed $J^P=1^-$.

At first sight, these two points should simplify the interference
picture in decays. However, decay properties may complicate the
situation. For example, the strong-interaction decays $J/\psi\to
\rho\pi$ go (up to smaller electromagnetic contributions) with
isospin conservation and are described by one coupling constant.
On the other side, weak decays $B^0\to\rho\pi$ (quark decay $\bar{b}
\to\bar{u}u\bar{d}$) violate isospin, and all three couplings of
$B^0$ to the three channels $\rho^0\pi^0, \,\,\rho^+\pi^-\,$ and
$\rho^-\pi^+\,$ may be independent. Experimentally, the branching
ratio for $\rho^0\pi^0$ is several times smaller than for
$\rho^\pm\pi^\mp$~\cite{be-r}. Difference of $CP$-violating
parameters for the decay channels $\rho^+\pi^-$ and  $\rho^-
\pi^+$~\cite{be-r} supports difference of the corresponding
decay amplitudes. Thus, all the charge channels look unrelated
indeed.

In addition, instead of the total energy, decays have another
variable, the time between production and decay of the hadron.
For neutral mesons with open flavor, the interference picture may
change with changing this time (see below).

Generally, decays of heavy hadrons provide a rich source of
resonances which are seen in multiparticle final states. If, e.g.,
we consider the three-particle decay $B^\pm\to\pi^+\pi^-\pi^\pm$,
its essential part comes from the quasi-two-body decay $\rho^0(770)
\pi^\pm$~\cite{ba-r}. But, in addition, there are also other
sub-decays: $f_0(980)\pi^\pm,\,\,f_2(1270)\pi^\pm$, and $\rho^0(1450)
\pi^\pm\,.$ This example demonstrates, that heavy hadron decays may
allow us to investigate various resonances insufficiently studied up
to now, including radial excitations. On the other side, the
presence of many resonances complicates the problem of their
accurate separation, because of numerous interference contributions.
Especially important (and difficult) is accounting for interference
between states with the same $J^P$-values, such as, e.g., $\rho(770)$
and $\rho(1450)$ or other radial excitations, since their
interference cannot be suppressed by angular integration of the
produced pion pair (just as in the case of direct interference).
Separation of such states may be done today only in a model-dependent
way. This is just what is done for $B^0$-meson decays~\cite{be-r}.

Sure, the above notes are qualitatively applicable to various
decays of $B$-mesons, as well as other heavy hadrons, e.g.,
of $D$-mesons, or charmed baryons. Interference is very interesting
also for final states with strange or even charmed hadrons. And, of
course, interference becomes even more essential for decays into
final states with four or more hadrons.

A specific feature of weak decays of hadrons is the possibility
of $CP$-violation. Manifestations of this phenomenon for neutral
flavored mesons, $K^0(\overline{K}^0),\,\, B_d^0(\overline{B}_d^0),
\,\,B_s^0(\overline{B}_s^0),\,$ and, possibly, $D^0(\overline{D}^0),
\,$ are closely related to interference between amplitudes of meson
and anti-meson decays. Such interference generates oscillatory
time behavior for particular decays of those mesons. The most
famous example is demonstrated by oscillations in the decay
$K^0\to2\pi$. But there exist less familiar manifestations
of interference also related to the $CP$-violation. Here we
briefly discuss two such effects.

An interesting problem is a discrete ambiguity in measuring
$CP$-violating parameters. Its origin may be traced~\cite{ars}
to a phase factor. Mathematically, this may be illustrated by a
simple example. Recall that any measurable value in quantum
physics is related to the absolute value squared of some amplitude.
Now, if one knows $|a|,\,\,|b|,\,$ and $$|c|^2=|a+b\cdot
e^{i\alpha}|^2\,,$$ one can determine the phase factor
$\exp(i\alpha)$ only with the two-fold ambiguity, up to the sign of
its imaginary part. This ambiguity will be eliminated, if one can
also find $$|c_\theta|^2=|a+ b\cdot e^{i(\alpha+\theta)}|^2\,,$$
where $\theta$ is a known function of some parameter and has a
definite (!) sign.

As an example, let us consider a particular weak decay $B^0
(\overline{B}^0)\to J/\psi K^0(\overline{K}^0)$, with the quark-level
decay $b\to c\overline{c}s$ or charge conjugate. The $CP$-violation
in this channel, as compared to any other $B$-decay, has been
measured most precisely (see recent overview~\cite{laz}). To resolve
ambiguity in this decay mode, it was suggested~\cite{az,ars} to study
the whole decay sequence, including the secondary kaon decay, in
dependence on both $t_B$ (time of the $B$-decay) and $t_K$ (time of
the $K$-decay). Then the coherent beauty-strangeness oscillations
provide the additional phase factor, which comes from the kaon time
evolution. It is related to the $(K_S,K_L)$ oscillations, and the
sign of its phase is determined by the known sign of the mass
difference $m_S-m_L\,$. Regretfully, the Monte Carlo
simulations~\cite{ars} show that such an approach needs
very high statistics, as yet unavailable.

More realistic has appeared another method, similar to that
suggested earlier~\cite{SQ} for the $B\to\rho\pi$ decays.
The BaBar Collaboration~\cite{ba-kp} investigated the decay
$$B\to J/\psi K\pi\,,$$ also with the quark-level decay $b\to
c\overline{c}s$. Here the reference sign for the $CP$-violating
phase comes from the interference of amplitudes for the
$(K\pi)$-pair produced in the $S$- and $P$-wave states. When the
$(K\pi)$ mass goes through the band of the resonance $K^*(890)$,
the $S$-wave phase stays nearly constant, while the $P$-wave phase
strongly changes (increases), according to the Breit-Wigner formula.
This known behavior of the phases (and of their difference in the
$S-P$ interference) has allowed experimentalists to eliminate the
sign ambiguity in the $CP$-violating phase factor as well~\cite{ba-kp}.

Another interesting (and somewhat unexpected) effect arises in decays
with secondary neutral kaons. It was first discovered theoretically
for decays $D\to K\pi$~\cite{az-DK}. The neutral kaons are usually
registered by their two-pion decay. It appears that $CP$-violating
$(K_S, K_L)$ interference in this secondary decay may imitate small
$CP$-violation for the $D$-meson branchings, even if it was totally
absent at the first stage of the decay. Such an effect was later
rediscovered, also theoretically, for the $\tau$-lepton decays
$\tau^\pm\to\nu\pi^\pm K_S$~\cite{BS}. Of course, this `secondary
violation' is totally determined by the neutral kaon properties and,
by itself, gives no new information. It should be present in any
decay with secondary neutral kaons, but, e.g., for the decay
$B^0(\overline{B}^0)\to J/\psi K_S$  it is practically unimportant
due to large $CP$-violation at the first step of the process. But
for $D$-meson or $\tau$-decays, with expected small $CP$-violation,
it may provide a useful reference point. Experimentally, this effect
has not yet been observed.

Even the few effects, briefly described here, demonstrate how
diverse may be interference manifestations in decays. They may be
very useful and important for studies of both spectroscopy and
properties of new resonances, as well as for more detailed
investigation of the known resonances.

\section{Discussion and conclusions}

Many examples of the interference of resonances, discussed in
this paper, are rather simple. They, nevertheless, allow us to
demonstrate various features, inherent also in more general and
complicated cases. That is why we are now able to formulate a
number of sufficiently general conclusions.

\begin{itemize}
\item{Interference of resonances has the same quantum nature as
      oscillations of particles, though they are observed in
      complementary variables - energy (mass) for the former,
      or time for the latter.}
\item{Two resonances can interfere even if they do not overlap,
      i.e., if their mass difference is large, larger than the
      sum of their widths. For instance, $\phi$ and $\omega$
      apparently interfere in several decay modes, though
      $M_\phi - M_\omega \approx 240~\textrm{MeV}$, while
      $\Gamma_\phi+\Gamma_\omega \approx 13~\textrm{MeV}$.
      Similarly, particle decays can reveal interference (and
      oscillations) even if lifetimes of the particles are
      essentially different. For instance, $K_S$ and $K_L$
      mesons demonstrate well-known oscillations, though
      $\tau_L/\tau_S\approx 500$.}
\item{If a resonance produces only a feeble signal (due to a rare
      decay mode, or due to mild production cross section), the
      contribution of its interference with a large background
      may appear more essential than the proper resonance signal.
      The corresponding background may be non-resonant, but may
      also come from another resonance having a profound signal.
      The both cases can be considered as amplification of the
      feeble resonance by the large background, be it resonance
      or non-resonance.}
\item{Interference of a resonance and a background may be either
      positive (constructive) or negative (destructive), depending
      on the relative phase between the resonance and the
      background. Moreover, the interference contribution usually
      has an additional energy dependence, in comparison with the
      familiar Breit-Wigner form, even if the background
      is energy-independent (it is more so for a resonant or
      any other energy-dependent background). Generally, the
      interference term changes its sign at some energy near
      the resonance position. }
\item{Generally, the interference reveals both bump and dip,
      with their positions shifted from the true position of
      the resonance. The relative intensity of the bump and
      the dip may be very different, essentially depending on
      the energy behavior of the background. Some cases may
      show only one kind of structure, either bump, or dip.}
\item{The same resonance may produce different interference
      pictures even in the same reaction when being observed in
      different decay modes. The situation is similar for particle
      oscillations: e.g., oscillations of neutral kaons
      look differently in semileptonic and two-pion decay channels.}
\item{Resonances can interfere in various ways. The simplest case
      is direct interference, when the resonances generate the same
      decay products. Evidently, this is possible only if (at least)
      some quantum numbers (such as flavors, baryon numbers, and so
      on) are the same for the interfering resonances. However,
      there can also be rescattering (or rearrangement) interference,
      when only some of the final particles may emerge in decays
      of both resonances. Such case of the resonance interference is
      more complicated. It needs correlated kinematics for products
      of the interfering resonances, but does not impose any
      restrictions on the resonance quantum numbers. Note that
      for the rescattering-type interference, the position of the
      interference bump (or dip) may, and even should, move when
      changing some parameters, e.g., momentum transfers.}
\item{Decays of heavy hadrons may demonstrate combinations of
      various kinds of interference. Account for these effects
      is necessary, and has been used, to separate different decay
      sub-channels, with different secondary resonances produced,
      and to extract related parameters. Regretfully, the structure
      of both the rescattering interference and different
      interference effects in decays is not yet clearly understood.
      That is why fits to experimental data are still very
      model-dependent in many cases.}

\end{itemize}

Concluding this brief discussion, one should emphasize that
direct interference has become a useful instrument for searching
and studying rare decays of well-established resonances. However,
its possibilities are limited by restrictions for the resonance
quantum numbers. Rescattering interference is not limited by
such requirements and, therefore, may provide effective methods
to search and study new resonances with arbitrary quantum numbers.
Data on multihadron decays of heavy particles also present a new
rapidly expanding area for applications of different kinds of
interference both to study spectroscopy of resonances and to
establish their characteristics.


\acknowledgments

The author thanks M~Amarian, D~Diakonov, K~Goeke, E~Pasyuk,
V~Petrov and M~V~Polyakov for useful discussions on the
interference problems. Very stimulating for the present paper were
questions at the meeting of the CLAS Hadron Spectroscopy Group and
at the seminars in Old Dominion University, Jefferson Laboratory,
and Ruhr University Bochum. Special thanks to I~Strakovsky, not
only for discussions, but also both for help in preparing this text,
and for its critical reading. Permissions of the KLOE Collaboration and
of the groups SND, CMD-2 allowed me to use their results for
illustration of the resonance peaks deformed in various ways by
interference effects. I also thank Professor K~Goeke for hospitality
in the Ruhr University Bochum, where a part of the text was written.
This work was supported in part by George Washington University,
by the US~Department of Energy under Grants DE--FG02--99ER41110 and
DE-FG02-96ER40960, and by Russian State Grant SS--3628.2008.2. The
author acknowledges partial support from Jefferson Lab, by the
Southeastern Universities Research Association under DOE contract
DE--AC05--84ER40150.


\end{document}